\documentclass[10pt,aps,twocolumn,floats,floatfix,amssymb,prd,superscriptaddress,nofootinbib]{revtex4-2}

\usepackage{graphicx}
\usepackage{amsmath,amssymb}
\usepackage{amsfonts}
\usepackage{xspace} 
\usepackage[usenames]{color}
\usepackage{dcolumn}
\usepackage{bm}
\usepackage{mathrsfs}
\usepackage[colorlinks=true]{hyperref}
\usepackage[all]{hypcap} 
\usepackage[utf8]{inputenc} 
\usepackage{multirow}
\usepackage{etoolbox}
\usepackage{tikz}

\usepackage[normalem]{ulem}

\begin{document}

\title{Shadows and Properties of Spin-Induced Scalarized Black Holes with and without a Ricci Coupling}

\author{Pedro G. S. Fernandes}
\affiliation{CP3-Origins, University of Southern Denmark, Campusvej 55, DK-5230 Odense M, Denmark}

\author{Clare Burrage}
\affiliation{School of Physics and Astronomy, University of Nottingham, University Park, Nottingham, NG7 2RD, United Kingdom}
\affiliation{Nottingham Centre of Gravity, University of Nottingham,
University Park, Nottingham NG7 2RD, United Kingdom}

\author{Astrid Eichhorn}
\affiliation{CP3-Origins, University of Southern Denmark, Campusvej 55, DK-5230 Odense M, Denmark}

\author{Thomas P. Sotiriou}
\affiliation{School of Physics and Astronomy, University of Nottingham, University Park, Nottingham, NG7 2RD, United Kingdom}
\affiliation{Nottingham Centre of Gravity, University of Nottingham,
University Park, Nottingham NG7 2RD, United Kingdom}
\affiliation{School of Mathematical Sciences, University of Nottingham, University Park, Nottingham NG7 2RD, United Kingdom}
\begin{abstract}
In this work, we explore the properties and shadows of spin-induced scalarized black holes, as well as investigate how a Ricci coupling influences them. Our findings reveal significant deviations from the Kerr metric in terms of the location and geodesic frequencies of the innermost stable circular orbit and light ring, with the former exhibiting more pronounced disparities. The shadows of scalarized black holes exhibit relatively minor deviations when compared to those of Kerr black holes with the same mass and spin. Overall, the presence of a Ricci coupling is observed to mitigate deviations from the Kerr metric.
\end{abstract}

\maketitle 

\section{Introduction}

In recent years, there has been significant interest in modified theories of gravity exhibiting black hole spontaneous scalarization \cite{Doneva:2017bvd,Silva:2017uqg,Dima:2020yac} (see Ref.~\cite{Doneva:2022ewd} for a recent review).
They align with General Relativity (GR) in the weak-field regime, where most gravity tests have so far been conducted \cite{Will:2014kxa}, while also permitting substantial deviations in the strong-gravity regime. These scenarios challenge the Kerr hypothesis \cite{Herdeiro:2022yle}, suggesting that in certain regimes, astrophysical black holes may not be described by the Kerr metric. Black hole spontaneous scalarization is often studied in scalar-Gauss-Bonnet gravity, where coupling a real scalar field $\phi$ to the Gauss-Bonnet invariant, $\mathcal{G} = R^2 - 4R_{\mu\nu}R^{\mu\nu} + R_{\mu\nu\alpha\beta}R^{\mu\nu\alpha\beta}$, induces a tachyonic instability around Kerr black holes under certain conditions. This coupling keeps the equations of motion at second-order and evades no-hair theorems \cite{Silva:2017uqg} (see also \cite{Herdeiro:2015waa,Sotiriou:2015pka} for reviews), thus allowing the unstable black hole to scalarize into a new stationary, non-Kerr geometry.

In the realm of scalar-Gauss-Bonnet theories, black hole scalarization takes on two distinct forms. 
When the Gauss-Bonnet coupling constant, $\alpha$, is positive, we encounter ``curvature-induced scalarization" \cite{Doneva:2017bvd,Antoniou:2017acq,Silva:2017uqg,Cunha:2019dwb}. In this case, the tachyonic instability is triggered by sufficiently high curvatures near the horizon. It is more pronounced in non-spinning black holes because the Gauss-Bonnet invariant is sign-definite for the Schwarzschild metric, while it can become zero and change sign outside the horizon for Kerr black holes with spins $J/M^2 \equiv j \geq 0.5$. Hence, close to the black hole's horizon, it can  contribute positively to the scalar's effective squared mass, counteracting the tachyonic behavior.
When the coupling $\alpha$ is negative, Kerr black holes with spins $j \geq 0.5$ become tachyonically unstable and  scalarize in a process known as ``spin-induced scalarization" \cite{Dima:2020yac,Herdeiro:2020wei,Berti:2020kgk,Doneva:2023oww,Doneva:2020nbb,Doneva:2020kfv}. 

The onset of scalarization is described well as a linear tachyonic instability and, hence, only interactions that are quadratic in the scalar contribute to it \cite{Andreou:2019ikc}. Nonetheless, the instability is quenched by nonlinearity, and hence the properties of its endpoint --- the scalarized black hole --- depend crucially on nonlinear interactions of the scalar \cite{Silva:2018qhn,Macedo:2019sem,Antoniou:2021zoy}. This means that additional couplings, which might not contribute to the theory linearized around a GR black hole, might  still determine the properties of scalarized black holes. A characteristic example is a coupling to the Ricci scalar \cite{Antoniou:2021zoy}.

We are motivated to incorporate such a coupling by several factors. First, one would expect it to be present from an effective field theory perspective as a lower-order coupling to curvature (in both derivatives and mass dimensions). Second, it plays a crucial role in making scalarization compatible with cosmological observations, by making GR a late-universe attractor \cite{Antoniou:2020nax}. Third, it improves the stability of scalarized black holes and mitigates the loss of hyperbolicity in radial perturbations \cite{Antoniou:2022agj}. Its positive effect on well-posedness at nonlinear level has also been demonstrated in the case of spherical collapse \cite{Thaalba:2023fmq}. Finally, it can inhibit scalarization in neutron stars \cite{Ventagli:2021ubn}, thereby alleviating potential strong constraints \cite{Danchev:2021tew}.

While the effect of the Ricci coupling on the charge and properties of spherical black holes that undergo ``curvature-induced" scalarization is fairly well understood \cite{Antoniou:2021zoy,Antoniou:2022agj,Thaalba:2023fmq}, its effect on black holes that are expected to develop hair through ``spin-induced" scalarization is unexplored. This is the case we want to study here. We will focus on a theory in which a scalar exhibits quadratic couplings to both the Gauss-Bonnet invariant and the Ricci scalar. Although this is not a complete effective field theory, these two coupling are the ones that break shift symmetry via coupling to curvature and are expected to have important contributions for scalarized black holes. Reflection symmetry $\phi \rightarrow -\phi$ can be invoked to remove the linear coupling of the scalar with curvature.

Our aim is to explore how spin-induced scalarization can affect observables, such as the scalar charge and black hole shadows, for black holes of different masses, and to uncover whether the Ricci coupling strengthens or suppresses deviations from the Kerr metric. To this end we generate numerical solutions to the field equations that describe rapidly spinning, scalarized black holes to high precision. We determine the properties and features of these solutions, including the scalar charge, the innermost stable circular orbit (ISCO), the photon ring and the shadow, and explore how they are affected by the size of the Ricci coupling and the size of the black holes with respect to the characteristic length scale associated with the Gauss-Bonnet coupling. 

\section{Theoretical Setup}

The theory under consideration in this work is defined by the following action:
\begin{equation}
    S = \frac{1}{16\pi} \int d^4 x \sqrt{-g} \left[ R - \left(\partial \phi\right)^2 + \phi^2\left( \frac{\alpha}{8} \mathcal{G} - \frac{\beta}{2} R \right) \right],
    \label{eq:action}
\end{equation}
where $\alpha$ has dimensions of length squared, $\beta$ is dimensionless, and we work in units $c=G=1$. Motivated by the results of Ref.~\cite{Antoniou:2020nax}, we take $\beta\geq 0$, because this sign ensures that GR solutions are cosmological attractors. The field equations resulting from varying the action in Eq.~\eqref{eq:action} with respect to the metric $g_{\mu \nu}$ and the scalar field $\phi$ are
\begin{equation}
    \begin{aligned}
        G_{\mu \nu} =& \partial_\mu \phi \partial_\nu \phi - \frac{1}{2} g_{\mu \nu} \left(\partial \phi\right)^2 + \frac{\alpha}{2} {}^{*}R^*_{\mu \alpha \nu \beta} \nabla^\alpha \nabla^\beta \phi^2 \\&  + \frac{\beta}{2} \left[G_{\mu \nu} + g_{\mu \nu} \Box - \nabla_\mu \nabla_\nu \right] \phi^2,
    \end{aligned}
    \label{eq:feqs}
\end{equation}
and
\begin{equation}
    \Box \phi = \left(\frac{\beta}{2} R - \frac{\alpha}{8} \mathcal{G} \right)\phi,
    \label{eq:sfeq}
\end{equation}
respectively, where ${}^{*}R^*_{\mu \alpha \nu \beta}$ denotes the double-dual of the Riemann tensor.

\begin{figure*}[]
	\centering
	\includegraphics[width=0.75\linewidth]{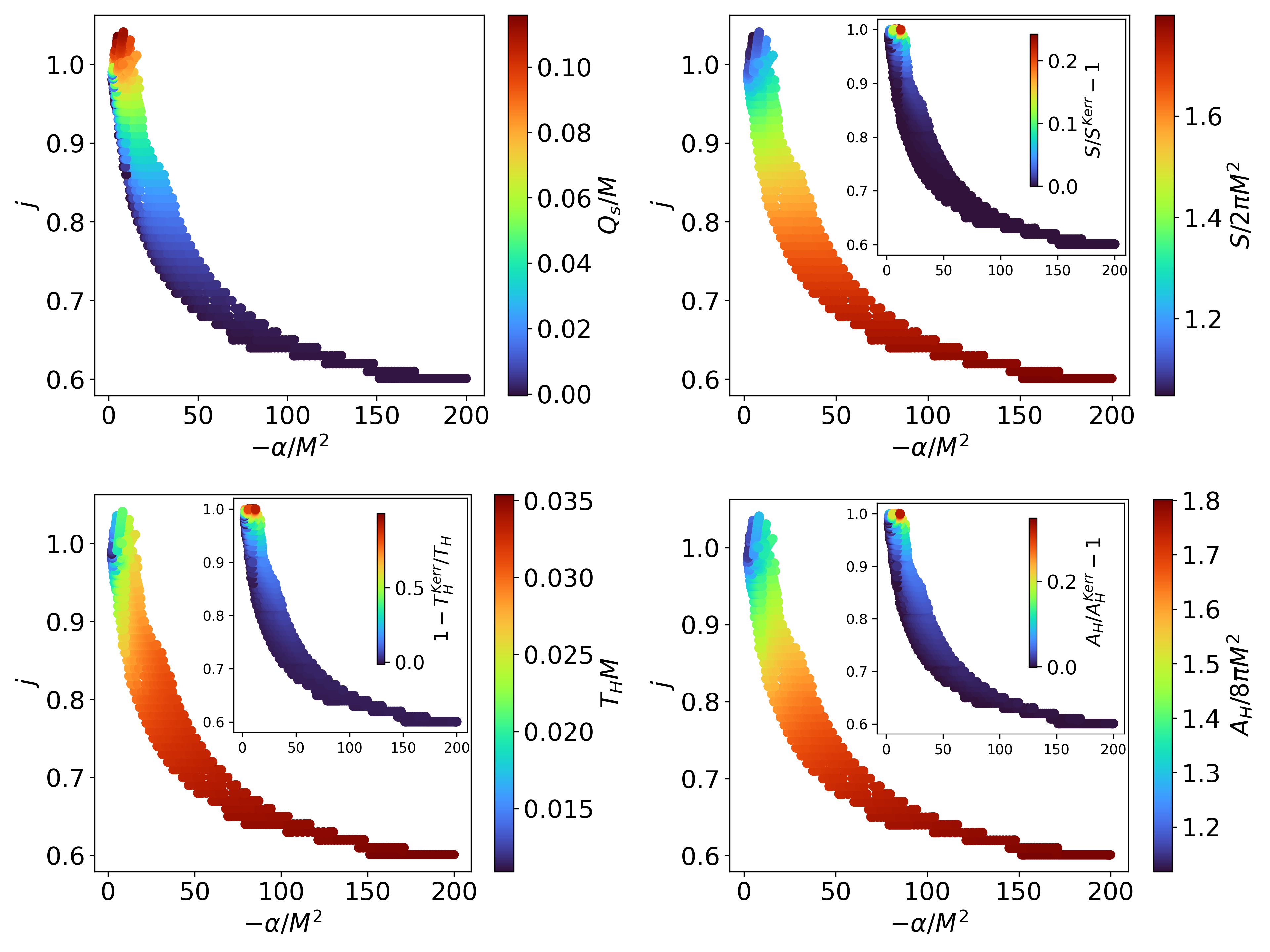}\hfill
     \caption{The plot depicts the domain of existence for scalarized black hole solutions ($\beta=0$) in the $(-\alpha/M^2, j)$ plane. Various quantities of interest are displayed as a heat map, organized from left to right as follows: scalar charge $Q_s$, entropy $S$ (top row); horizon temperature $T_H$, horizon area $A_H$ (bottom row). These quantities are properly normalized relative to the black hole's mass. In general, deviations from the Kerr metric are larger for larger spins and coupling strengths.}
	\label{fig:domain}
\end{figure*}

\begin{figure*}[]
	\centering
	\includegraphics[width=0.75\linewidth]{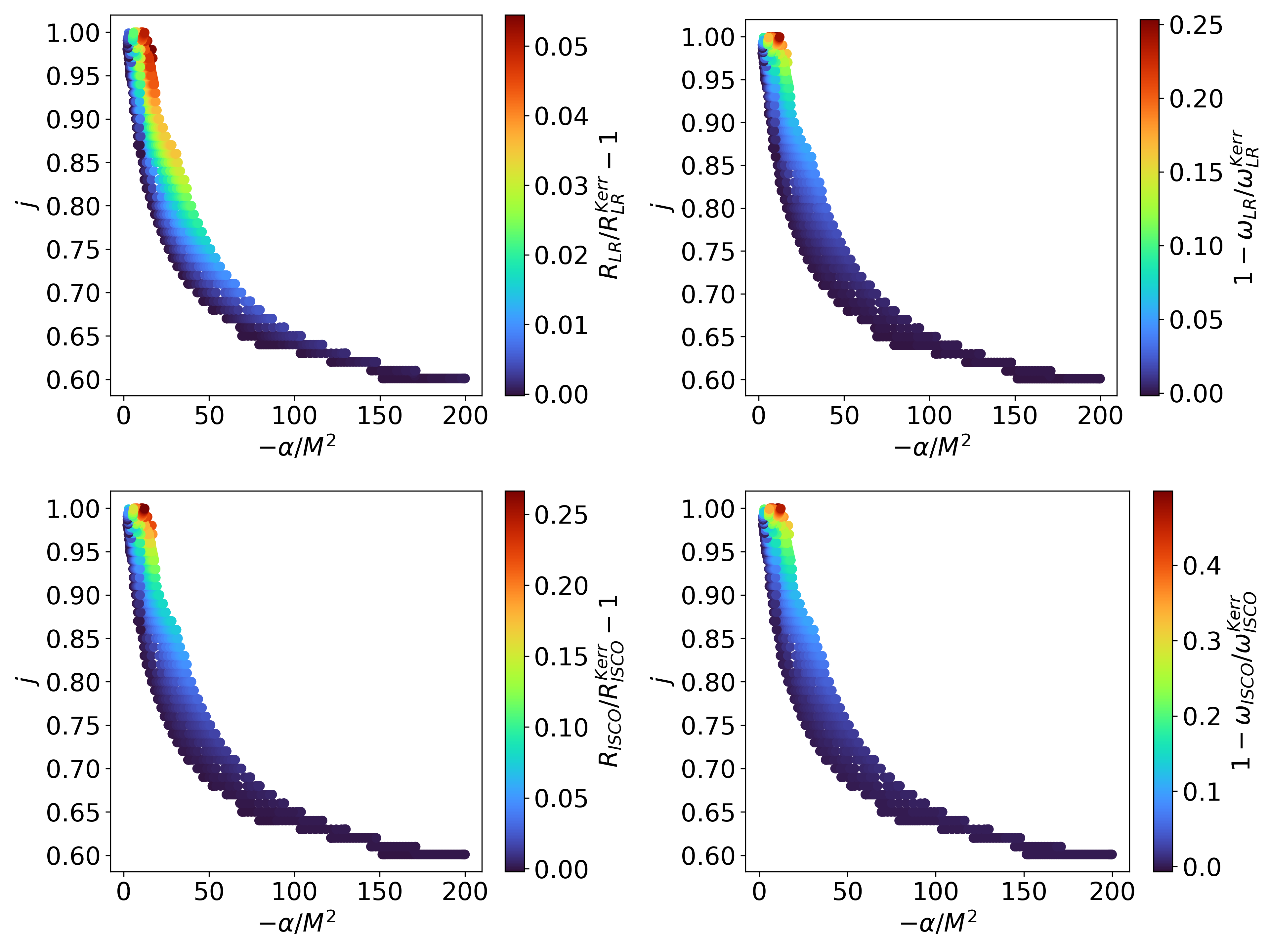}\hfill
     \caption{Domain of existence for scalarized black hole solutions ($\beta=0$) in the $(-\alpha/M^2, j)$ plane. Various quantities of interest are displayed as a heat map, organized from left to right as follows: fractional deviation from Kerr in the location of the light ring, $\mathfrak{R}_{LR}/\mathfrak{R}_{LR}^{Kerr}-1$, fractional deviation from Kerr in the geodesic frequency at the light ring, $1-\omega_{LR}/\omega_{LR}^{Kerr}$ (top row); fractional deviation from Kerr in the location of the ISCO, $\mathfrak{R}_{ISCO}/\mathfrak{R}_{ISCO}^{Kerr}-1$, and fractional deviation from Kerr in the geodesic frequency at the ISCO, $1-\omega_{ISCO}/\omega_{ISCO}^{Kerr}$ (bottom row). These quantities are properly normalized relative to the black hole's mass. In general, deviations from the Kerr metric are larger for larger spins and coupling strengths. Both the ISCO and light ring of scalarized black holes are positioned farther from the black hole's horizon compared to a Kerr black hole, and their geodesic frequencies are correspondingly lower. The disparities are more pronounced in the case of the ISCO.}
	\label{fig:domain2}
\end{figure*}

Metrics that are solutions to Einstein's equations in vacuum together with $\phi=0$ are admissible solutions to Eqs.~\eqref{eq:feqs} and \eqref{eq:sfeq}. However, these solutions are not necessarily stable.
Examining the scalar field equation \eqref{eq:sfeq}, we can discern that the combination $\frac{\beta}{2}R - \frac{\alpha}{8}\mathcal{G}$ effectively acts as a squared mass term for the scalar field $\phi$ and its perturbations. As we are interested purely in spin-induced scalarization, we take $\alpha$ to be negative from this point onward.
Then, scalar perturbations around a Kerr black hole tend to become tachyonic when the black hole's spin exceeds a certain threshold, specifically, $j\geq 0.5$ \cite{Dima:2020yac}. 
This occurs because the Gauss-Bonnet invariant of the Kerr geometry no longer maintains a definite sign for high enough spins and the combination $-\alpha \mathcal{G}$, entering the effective squared mass for the scalar, becomes negative in certain regions close to the horizon, while $R=0$. As demonstrated in Refs.~\cite{Herdeiro:2020wei,Berti:2020kgk}, this scenario gives rise to non-trivial scalarized black holes when the coupling constant $\alpha$ is sufficiently negative. In particular, Ref.~\cite{Herdeiro:2020wei} considered a quadratic exponential coupling of the scalar to the Gauss-Bonnet invariant, while Ref.~\cite{Berti:2020kgk} considered a simple quadratic coupling, as we do in this work. In both cases, scalarized black holes were observed to be entropically favoured over Kerr black holes with the same mass and spin.

\section{Numerical Procedure}

To construct the stationary and axially symmetric black-hole solutions to the field equations \eqref{eq:feqs} and \eqref{eq:sfeq}, we will follow the approach of Ref.~\cite{Fernandes:2022gde}, using a publicly available code developed by one of us \cite{gitlink}. This code and approach have been previously used with success, e.g., in Refs.~\cite{Fernandes:2022gde,Burrage:2023zvk,Lai:2023gwe,Guo:2023mda,Xiong:2023bpl,Eichhorn:2023iab}. These solutions possess two commuting Killing vector fields, $\xi = \partial_t$ and $\eta = \partial_\varphi$, in an adapted coordinate system. 
We assume that the metric is circular, such that an ansatz in terms of four free, dimensionless functions $f,g,h,W$ of $r$, $\theta$ in quasi-isotropic coordinates is adequate \cite{Papapetrou:1966zz}
\begin{equation}
\begin{aligned}
    ds^2 =& -f \left(1-\frac{r_H}{r}\right)^2 dt^2 + \frac{g h}{f} \left(dr^2 + r^2 d\theta^2\right),\\& + \frac{g}{f} r^2 \sin^2\theta \left(d\varphi - \frac{W r_H}{r^2} dt\right)^2.
\end{aligned}
\label{eq:metric}
\end{equation}
Here, $r_H$ is the coordinate location of the event horizon. In our numerical setup, we employ the compactified radial coordinate $x=1-2r_H/r$, mapping the interval $[r_H, +\infty)$ to $[-1, 1]$. We will consider only the case of an even parity scalar field, $\phi(r,\theta-\pi) = \phi(r,\theta)$. Therefore all functions have definite (even) parity with respect to $\theta=\pi/2$, and we consider only the range $\theta \in [0, \pi/2]$ in our numerical setup.

We employ boundary conditions as follows: Regularity, axial symmetry and parity considerations imply $\partial_\theta f = \partial_\theta g = \partial_\theta h = \partial_\theta W = \partial_\theta \phi = 0$, at $\theta = 0, \, \pi/2$. At the horizon ($x=-1$) our functions obey $f-2 \partial_x f = g+2 \partial_x g = \partial_x h = W - r_H \Omega_H = \partial_x \phi = 0$, where $\Omega_H$ is the angular velocity of the horizon. Asymptotic flatness is imposed by the following boundary conditions at $x=1$: $f=g=h=1$, and $W = \phi = 0$. The angular momentum $J$ and the Arnowitt-Deser-Misner (ADM) mass $M$ can be extracted from the asymptotic fall-offs of the metric functions: $g_{tt} \sim -1 + 2M/r + \mathcal{O}\left(r^{-2}\right), \quad g_{\varphi t} \sim 2J\sin^2 \theta/r^2 + \mathcal{O}\left(r^{-3}\right)$. The scalar decays as $\phi \sim Q_s/r + \mathcal{O}\left(r^{-2}\right)$, where $Q_s$ is the scalar charge of the solution. Again, we define $j\equiv J/M^2$.

To solve the partial differential equations resulting from the field equations, we have used the code described in Ref.~\cite{Fernandes:2022gde}, which employs a pseudospectral method together with the Newton-Raphson root-finding algorithm to solve the non-linear system (see also Ref.~\cite{Dias:2015nua}). We expand each of the functions in a spectral series with resolution $N_x$ and $N_\theta$ in the radial and angular coordinates $x$ and $\theta$, respectively. The spectral series we use for each of the functions $\mathcal{F}^{(k)}=\{f,g,h,W,\phi\}$ is given by
\begin{equation}
  \mathcal{F}^{(k)} = \sum_{i=0}^{N_x-1} \sum_{j=0}^{N_\theta-1} c_{ij}^{(k)} T_i(x) \cos \left(2j\theta\right),
\label{eq:spectralexpansion1}
\end{equation}
where $T_i(x)$ denotes the $i^{th}$ Chebyshev polynomial, and $c_{ij}^{(k)}$ are the spectral coefficients. Note that the angular boundary conditions are automatically satisfied with this spectral expansion, and need not be explicitly imposed in the numerical method.

In our setup, we have three input parameters: $(r_H, \Omega_H, \alpha)$. We use a resolution of $N_x \times N_\theta = 40 \times 8$ in most cases, and a higher resolution $N_x \times N_\theta = 42 \times 12$ for very large spins.
We can estimate the numerical error in our solutions using a Smarr-type relation given below in Eq.~\eqref{eq:smarr}.
The estimated error is typically less than $\mathcal{O}\left(10^{-8}\right)$, although errors increase for large spins, $j$, and/or coupling strength. We have only accepted an output of our solver if the error as estimated by the Smarr-type relation is $\mathcal{O}\left(10^{-5}\right)$ or less.

\section{Results}
\subsection{Physical properties}
In this section we examine some physical properties of the scalarized black holes, namely the scalar charge $Q_s$, Hawking temperature $T_H$ \cite{Hawking:1975vcx}, event horizon area $A_H$, entropy $S$, and the location of the co-rotating light ring and innermost stable circular orbit (ISCO) and respective geodesic frequencies.
The Hawking temperature $T_H$ \cite{Hawking:1975vcx} and event horizon area $A_H$ are given by
\begin{equation}
  T_H = \frac{1}{2\pi r_H} \frac{f}{\sqrt{g h}}, \quad A_H = 2\pi r_H^2 \int_{0}^{\pi} d\theta \sin \theta \frac{g \sqrt{h}}{f},
\end{equation}
where the above expressions are to be evaluated at the horizon. The entropy of the black hole does not follow the Bekenstein-Hawking relation, but can be expressed as an integral over the horizon
\begin{equation}
    S = \frac{1}{4} \int_\mathcal{H} d^2x \sqrt{\gamma} \left[ 1 - \frac{\beta}{2} \phi^2 + \frac{\alpha}{4}\phi^2 \Tilde{R} \right],
    \label{eq:entropy}
\end{equation}
where $\gamma$ is the determinant of the induced metric on the horizon and $\Tilde{R}$ its Ricci scalar.
For stationary and axially-symmetric
black hole solutions of the theory in Eq.~\eqref{eq:action}, 
the Smarr-type
relation is \cite{Fernandes:2022gde,Cunha:2019dwb}
\begin{equation}
    M = 2 T_H S + 2\Omega_H J +\frac{1}{8\pi} \int d^3x \sqrt{-g} \phi\left[\Box - \frac{\beta}{2}R\right] \phi.
    \label{eq:smarr}
\end{equation}

In Fig.~\ref{fig:domain}, the domain of existence for spin-induced scalarized black holes without a Ricci coupling ($\beta=0$) is depicted in the $(-\alpha/M^2,j)$ plane.
Various relevant physical properties are represented as a heat map. The plots display the scalar charge $Q_s$, entropy $S$, Hawking temperature $T_H$, and the horizon area $A_H$ of these black holes. In the insets within these plots, a comparative analysis is presented between each of these properties for scalarized black holes and Kerr black holes, when non-uniqueness holds. It is worth noting that studies of these four properties in related models have previously been conducted in Refs.~\cite{Herdeiro:2020wei,Berti:2020kgk} and provide an additional, independent validation of our numerical results. In specific regions of the parameter space, we have identified solutions with spins reaching up to $j\approx 1.04$. In contrast, Refs.~\cite{Herdeiro:2020wei,Berti:2020kgk} have documented solutions with maximum spins of approximately $j\approx 1.01$. We suspect our numerical method was able to find these highly-spinning solutions due to the use of a spectral method.\footnote{It is a critical open question whether astrophysical black holes can access this high-spin regime. In GR, dynamical limits on the spin lie below $j=1$, but this may well be different in modified theories. If this is the case, observational constraints on black-hole spins \cite{Reynolds:2020jwt} may become of interest in probing scalarized black holes.}

\begin{figure}[]
	\centering
	\includegraphics[width=\linewidth]{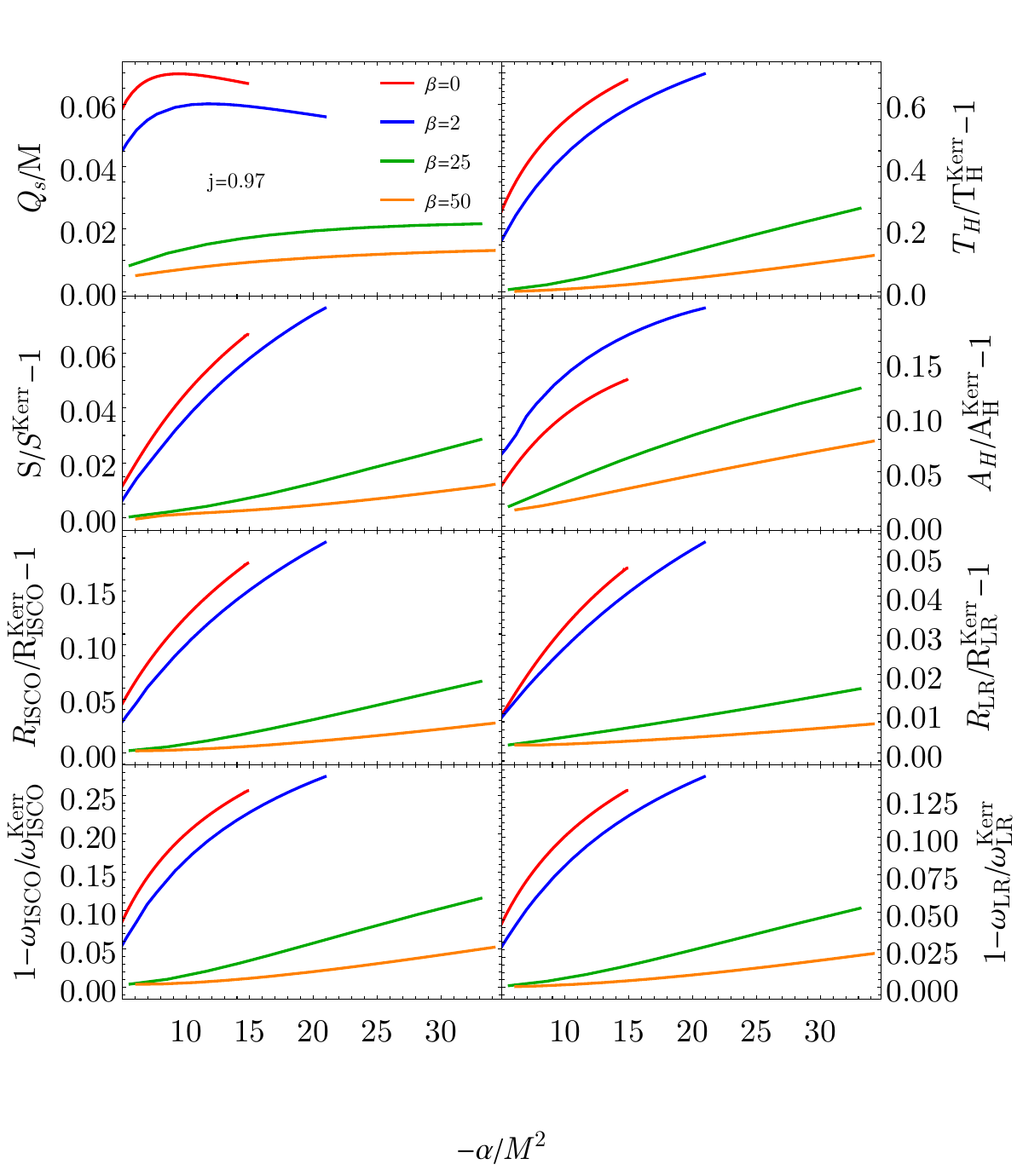}\vfill
        \caption{Impact of a Ricci coupling on physical quantities of interest for scalarized black holes with $j=0.97$. There are, organized from top to bottom as follows: scalar charge $Q_s$, entropy $S$, location of the ISCO $\mathfrak{R}_{ISCO}$, geodesic frequency at the ISCO $\omega_{ISCO}$ (left column); horizon temperature $T_H$, horizon area $A_H$, location of the light ring $\mathfrak{R}_{LR}$, and geodesic frequency at the light ring, $\omega_{ISCO}$ (right column).}
	\label{fig:BetaCompare}
\end{figure}

\begin{figure*}[]
	\centering
	\includegraphics[width=\linewidth]{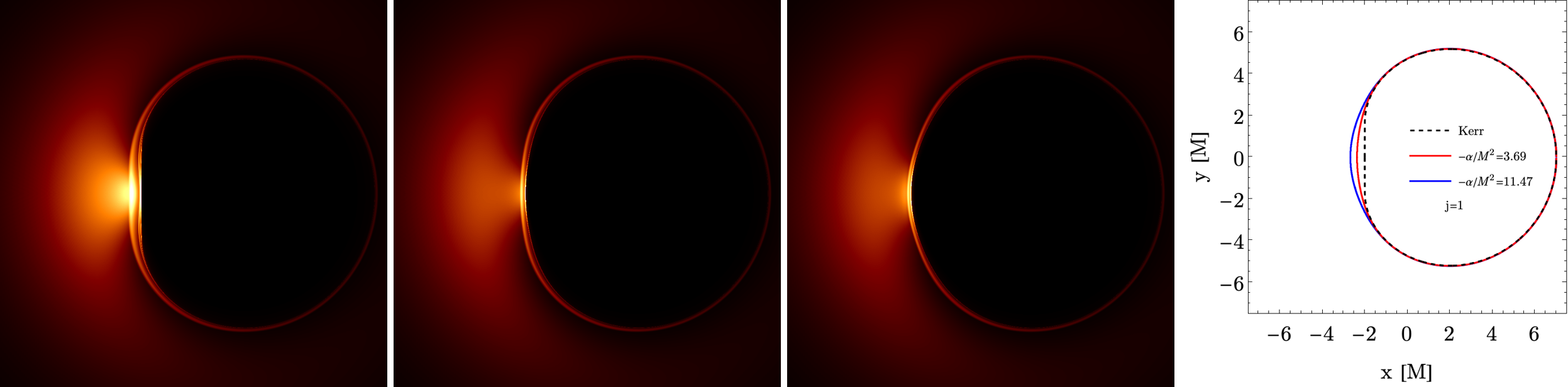}\hfill
	\caption{Simulated images of black holes with $j=1$, for the theory with $\beta=0$, using the emission model of Refs.~\cite{Cardenas-Avendano:2022csp, Gralla:2020srx}. From left to right: a Kerr black hole, scalarized black holes with $-\alpha/M^2\approx 3.69$ and $-\alpha/M^2\approx 11.47$, and a comparison of their shadow boundaries.}
	\label{fig:shadows}
\end{figure*}

In Fig.~\ref{fig:domain2}, the plots pertain to the marginally stable circular orbits for both massive and massless particles, the ISCO and the light rings. The ISCO represents the smallest possible
radius for a stable circular orbit of massive particles and is often considered as the inner boundary of an accretion disk encircling a black hole. Charged particles in accelerated motion around the black hole emit synchrotron radiation, and the characteristics of this radiation are associated with the geodesic frequencies at the ISCO. Consequently, by examining the ISCO through accretion disks, it is possible to deduce various physical attributes of an astrophysical black hole \cite{Abramowicz:2011xu}. Light rings are circular null geodesics, usually unstable, where light can revolve around a black hole before getting dispersed towards infinity or entering the event horizon. From an observational perspective, these light rings hold significance in observations conducted by the Event Horizon Telescope collaboration \cite{EventHorizonTelescope:2019ggy,EventHorizonTelescope:2020qrl,EventHorizonTelescope:2021dqv,EventHorizonTelescope:2022xqj}, given their close association with the black hole's shadow. The angular frequency at the light ring is related to the timescale of the black hole's response when subjected to perturbations. The physical quantities related to the light rings and the ISCO, namely the location and geodesic frequency, were computed in this work following Ref. \cite{Fernandes:2022gde}.

The first column plots in Fig.~\ref{fig:domain2} compare the perimetral radius $\mathfrak{R}=\sqrt{g_{\varphi \varphi}}|_{\theta=\pi/2}$ of the co-rotating ISCO and light ring to that of a Kerr black hole. The perimetral radius $\mathfrak{R}$ is a geometrically meaningful coordinate defined such that the circumference of a circle along the equatorial plane is $2\pi \mathfrak{R}$. The comparisons are made in the region of non-uniqueness.
Differences from the Kerr metric increase with higher spin and coupling strength. The maximum difference in the perimetral location of the light ring is $\mathfrak{R}_{LR}/\mathfrak{R}_{LR}^{Kerr}-1 \sim \mathcal{O}\left( 5\%\right)$, while the maximum difference for the ISCO perimetral location exceeds $\mathfrak{R}_{ISCO}/\mathfrak{R}_{ISCO}^{Kerr}-1 \sim \mathcal{O}\left( 25\%\right)$.
The second column of plots in Fig.~\ref{fig:domain2} presents a comparison of the geodesic frequencies $\omega$ between the co-rotating ISCO and light ring of scalarized black holes and those of a Kerr black hole. The most significant difference is approximately $1-\omega_{LR}/\omega_{LR}^{Kerr} \sim \mathcal{O}\left( 25\%\right)$, while the deviations for the ISCO are roughly $1-\omega_{ISCO}/\omega_{ISCO}^{Kerr} \sim \mathcal{O}\left( 50\%\right)$. 
Generically, scalarized black holes are less compact than Kerr black holes of the same ADM mass. This is manifested by the event horizon having larger area and the ISCO and light rings lying at larger radii (and the corresponding frequencies being lower, i.e., the time scales larger).

We now examine how a Ricci coupling influences the properties of scalarized black holes. Our focus on a Ricci coupling stems from its potential to scalarize larger black holes with respect to conventional black-hole scalarization models, without conflicting with observational data. This is achieved by inhibiting scalarization in neutron stars \cite{Ventagli:2021ubn}, thereby complementing its other advantageous features \cite{Antoniou:2022agj,Antoniou:2020nax,Thaalba:2023fmq}.
To demonstrate our findings, we present a series of scalarized black holes with a relatively high spin, specifically $j=0.97$, for various values of $\beta=\{0,2,25,50\}$, as shown in Fig.~\ref{fig:BetaCompare}. Our qualitative observations hold true for other spin values that we have examined.
At larger Ricci couplings physical deviations from the Kerr geometry are smaller, and the scalar charge is suppressed.

Heuristically, a large Ricci coupling in the action requires solutions to have smaller Ricci curvature and thus be closer to the GR vacuum solution with vanishing Ricci curvature, i.e., a Kerr black hole. However, there is an exception to this general trend concerning the event horizon area, $A_H$. For values of $\beta$ around $\sim \mathcal{O}(1)$, we actually observe greater deviations compared to Kerr. This pattern shifts when even higher values of the Ricci coupling are taken into account, as deviations in $A_H$ then start to diminish. Our hypothesis is that the rise in $A_H$ is a response to offset the negative contribution introduced by the Ricci coupling on the entropy of scalarized black holes, as indicated in Eq.~\eqref{eq:entropy}, meaning that the entropy of a scalarized black hole exceeds that of a Kerr black hole with equivalent mass and spin.
Also, higher values of the Ricci coupling widen the domain of existence of solutions to higher values of the Gauss-Bonnet coupling, $\alpha$.

\subsection{Shadows}

Lastly, we explore shadows cast by spin-induced scalarized black holes. Before we present our results a few comments on the size of $\alpha$ are due. One can scale the coordinates by the black hole mass to render the equations dimensionless (in units where $G=c=1$). $\alpha$ is then replaced with its dimensionless version $\alpha/M^2$ ($\beta$ is dimensionless). This is the approach we have followed in the previous section, and hence our results apply to any black hole mass. It is clear from these results however, that highly spinning scalarized black holes of appreciable charge only exist when $\alpha/M^2$ is close to 1. For much smaller values, the tachyonic instability would not be triggered unless rotation is unrealistically close to the maximal value \cite{Dima:2020yac} and for larger values, scalarized black holes only exist for low spins and carry small charges. Hence, for supermassive black holes, for which shadow observations are possible through Very-Long-Baseline-Interferometry (VLBI) experiments \cite{Ayzenberg:2023hfw,EventHorizonTelescope:2019ggy, EventHorizonTelescope:2020qrl, EventHorizonTelescope:2021dqv, EventHorizonTelescope:2022xqj}, one would need $\sqrt{|\alpha|} \gtrsim \mathcal{O}\left(10^5 M_\odot\right)$.

There are two obvious issues with such a large $\alpha$. First, there is the effect it would have on neutron stars, which can exhibit a tachyonic instability for both signs of $\alpha$ \cite{Silva:2017uqg,Ventagli:2020rnx}. A value of $\alpha$ that would allow for supermassive black hole scalarization could either lead to neutron star scalarization levels that are incompatible with observations \cite{Danchev:2021tew,Ventagli:2021ubn} or simply render the GR solutions unstable without a scalarized counterpart \cite{Ventagli:2021ubn}. The second issue is that black holes have a minimum size in the model we are considering. A large $\alpha$ would mean that the minimum size of black holes is far larger than a few solar masses, requiring the rather speculative assumption that gravitational-wave signals observed by the LIGO/VIRGO/KAGRA collaboration are in fact generated by the mergers of some exotic compact objects. The inclusion of the Ricci coupling can in principle quench neutron star scalarization and reduce the minimum mass of black holes, but making both 
effects strong enough
for such large values of $\alpha$ is likely to require large values of $\beta$ as well. The results of the previous section (as well as those of \cite{Antoniou:2021zoy}) show that this would in turn suppress the scalar charge for black holes.

With the above in mind, we study shadows for three distinct reasons: first, to demonstrate explicitly that large values of $\beta$ makes shadows virtually indistinguishable from those of Kerr black holes; second, because one can think of shadows as theoretical probes of the geometry irrespective of their mass; and third, because there might be scalarization models in which supermassive black holes get selectively scalarized, without neutron stars exhibiting a tachyonic instability. An example of such a model for curvature-induced scalarization was recently presented in Ref.~\cite{Eichhorn:2023iab}.

\begin{figure}[]
	\centering
	\includegraphics[width=\linewidth]{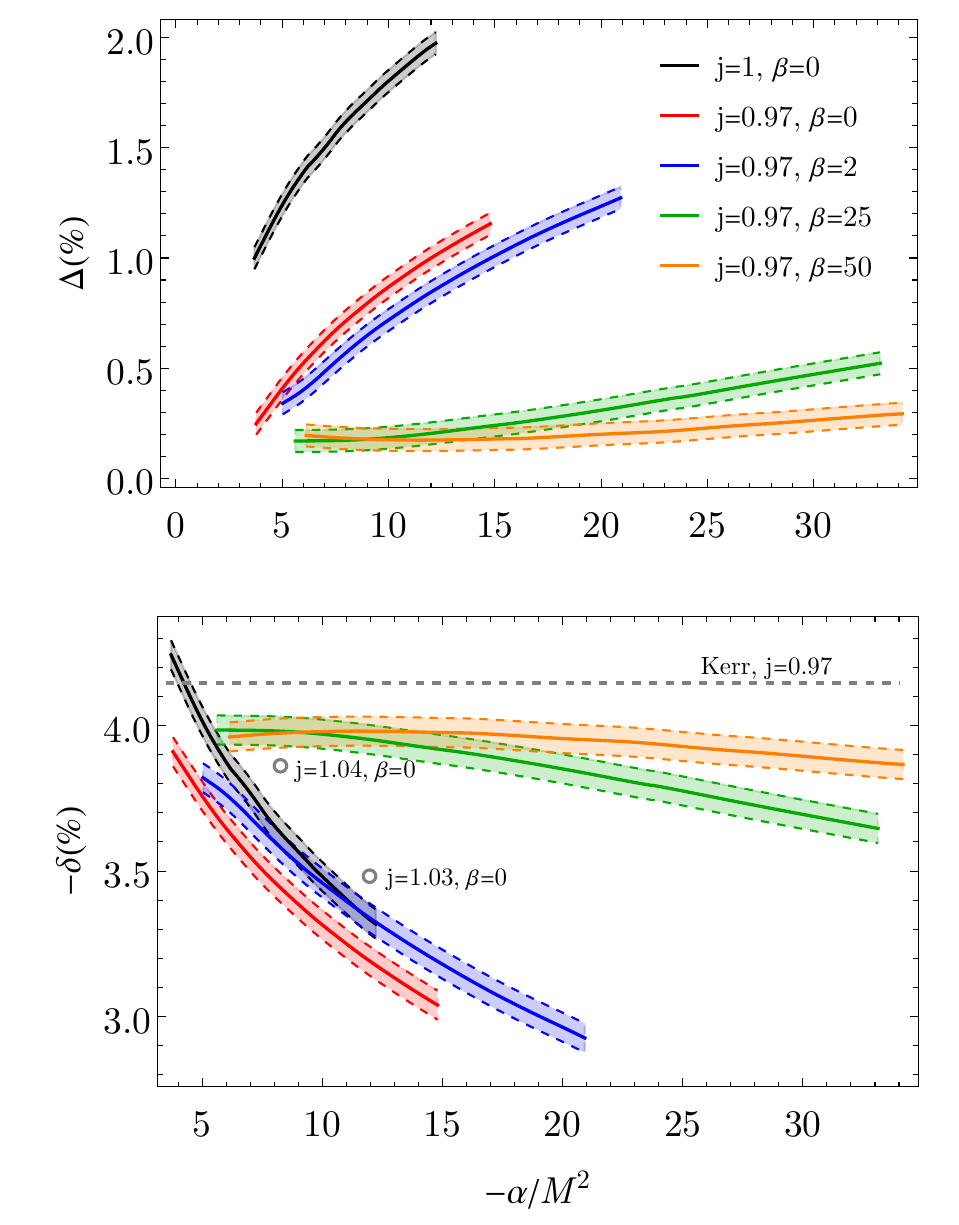}\vfill
	\caption{Shadow deviation parameters $\Delta$ and $\delta$ for a set of highly spinning black holes in theories with different values of Ricci coupling. The gray points refer to single overspinning solutions. The error bands include an estimated numerical error of $\pm 0.05\%$.} 
	\label{fig:deltashadows}
\end{figure}

To generate simulated images of these black holes, we employ the publicly available code \emph{FOORT} \cite{FOORT, Staelens:2023jgr} for the backward tracing of null geodesics. In our setup, we positioned an observer off-center on the equatorial plane, $\theta=\pi/2$, because at this inclination we anticipate the largest deviations from the Kerr geometry. The observer is situated at a distance of $r = 1000M$, with a $15M \times 15M$ viewscreen. We tracked a total of $1024 \times 1024$ trajectories, following them until they either vanish into the horizon or escape beyond the celestial sphere, which is also positioned at $r=1000M$.
Additionally, we monitor the number of times a geodesic crosses the equatorial plane to construct an emission profile, following the model outlined in Refs.~\cite{Cardenas-Avendano:2022csp, Gralla:2020srx}. This approach enables us to replicate realistic images\footnote{We have used the parameters $\mu=1$, $\gamma=-1.5$, $\sigma=3$, $\xi=\beta_r=\beta_\phi=1$.}.

\begin{figure}[]
	\centering
	\includegraphics[width=\columnwidth]{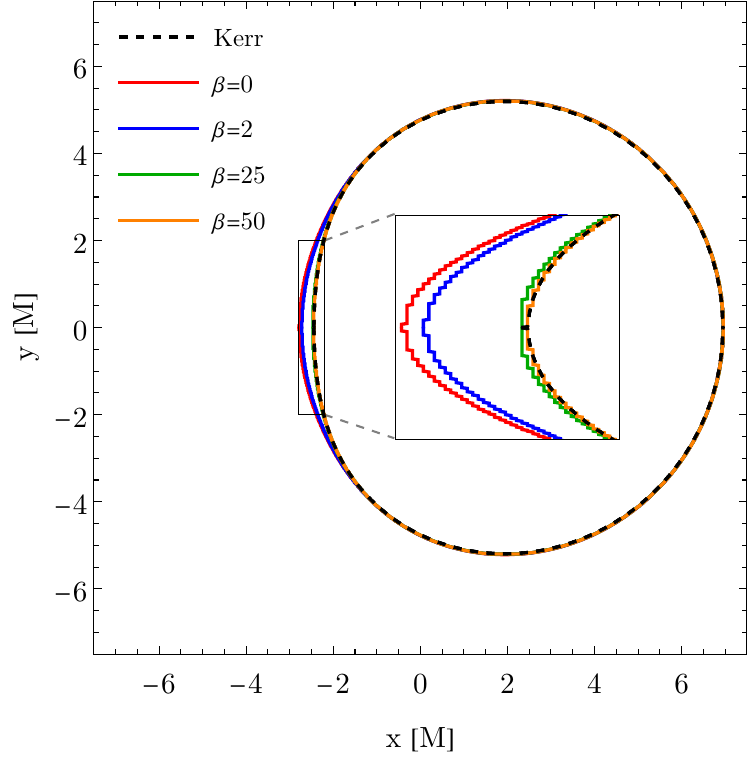}\hfill
 	\includegraphics[width=\columnwidth]{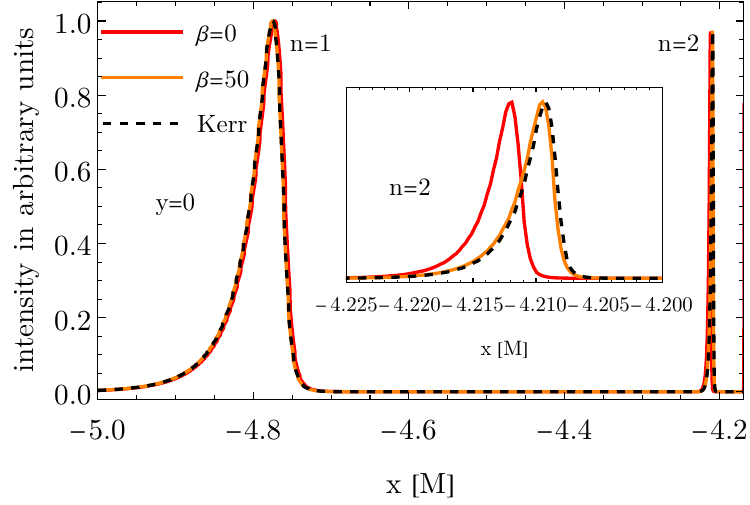}
    \caption{Impact of the Ricci coupling on the shadow boundary and photon rings ($n=1$ and $n=2$) of scalarized black holes with $j=0.97$ and $-\alpha/M^2\approx 13$.}
	\label{fig:shadowsbeta}
\end{figure}

A proper connection to observational data would require a simulated observation and a fit to the simulated data in the Fourier-plane of the image, while marginalizing over disk models. Here, we instead only estimate whether or not such a study would be likely to result in detectable deviations from the Kerr geometry, given current observational capabilities. To do so, we will employ a straightforward measure of the size of the black hole shadow: the areal radius, denoted as $\mathcal{R}$. The areal radius is determined as the radius of a circle with the same area $\mathcal{A}$ as the shadow, mathematically expressed as $\mathcal{R} = \sqrt{\mathcal{A}/\pi}$. This parameter is directly linked to the measured angular size of the black-hole shadow, and is well defined even for non-circular shadows.
Using the areal radius, we introduce two dimensionless quantities to characterize the shadow of a scalarized black hole: the Kerr shadow deviation $\Delta$, and the Schwarzschild shadow deviation $\delta$. These are defined as:
\begin{equation}
    \Delta = \frac{\mathcal{R}_{\mbox{scalarized}}}{\mathcal{R}_{\mbox{Kerr}}}-1, \quad \delta = \frac{\mathcal{R}_{\mbox{scalarized}}}{\mathcal{R}_{\mbox{Schwarzschild}}}-1.
\end{equation}
where $\mathcal{R}_{\mbox{scalarized}}$ is the areal radius of the scalarized black-hole shadow, $\mathcal{R}_{\mbox{Kerr}}$ is the areal radius for a Kerr black hole with the same mass and spin, and $\mathcal{R}_{\mbox{Schwarzschild}} = 3\sqrt{3}M$.
The Kerr deviation $\Delta$ compares the shadow radius of a scalarized black hole to the shadow of a Kerr black hole with the same mass and spin, and therefore it is only defined for spins $j \leq 1$. The Schwarzschild deviation $\delta$ is defined for any spin value. In particular, from current observations, we have \cite{EventHorizonTelescope:2019ggy, EventHorizonTelescope:2020qrl, EventHorizonTelescope:2021dqv, EventHorizonTelescope:2022xqj, Vagnozzi:2022moj}
\begin{equation}
\begin{aligned}
    & -18\% \lesssim \delta \left(\mbox{M87}\right) \lesssim 16\%, \quad (1\sigma)\\&
    -12.5\% \lesssim \delta \left(\mbox{Sgr A}^*\right) \lesssim 0.5\%. \quad (1\sigma)
\end{aligned}
\label{eq:shadowbounds}
\end{equation}

In Fig.~\ref{fig:shadows}, we display simulated images of scalarized black holes with $j=1$ and no Ricci coupling, comparing them with those of a Kerr black hole. We observe that as we increase the coupling strength, the shadow boundary of the scalarized black hole gradually transitions from the characteristic ``D-shape" associated with extremal Kerr black holes to a more circular form. However, in terms of shadow area, the differences remain relatively minor, as illustrated in Fig.~\ref{fig:deltashadows}. For all the examined shadows, even those corresponding to black holes with spins exceeding $j>1$, they fall well within the boundaries in Eq.~\eqref{eq:shadowbounds}. The largest deviation in $\delta$ is of the order of approximately $\mathcal{O}\left(-4\%\right)$. 

We have further investigated the influence of a Ricci coupling on the shadows of scalarized black holes. We find the same trend we observed previously in other physical parameters, which is that as the Ricci coupling increases, the shadow becomes progressively more similar to that of a Kerr black hole. This trend is visually depicted in both Figs.~\ref{fig:deltashadows} and \ref{fig:shadowsbeta}, where we focus on black holes with a spin of $j=0.97$. In these figures, it becomes evident that for $\beta = 50$, the distinctions between the shadow boundaries of scalarized black holes and Kerr black holes are practically negligible. Similar conclusions hold for other observables, such as the photon rings as shown in Fig.~\ref{fig:shadowsbeta} (bottom). It is worth to comment on photon rings, given that they are observables that are rather clean probes of the spacetime geometry, and -- unlike many other image features -- less contaminated by astrophysical effects. We find that a scalarized black hole generically has a smaller separation between subsequent photon rings than its Kerr counterpart, cf.~Fig.~\ref{fig:shadowsbeta}. If this is a generic feature, holding also, e.g., in the model in \cite{Eichhorn:2023iab}, even higher resolution is required to separate photon rings than in the case of the Kerr spacetime, posing an added challenge for VLBI observations.

\section{Conclusions}
In this study, we have investigated spin-induced scalarization of black holes triggered by a quadratic coupling of a scalar to the Gauss-Bonnet invariant. We have focused on how properties of such black holes, such as entropy, horizon area, temperature, and observables, such as the ISCO, and shadows, deviate from their Kerr black hole counterparts for different values of the couplings. We have also explored for the first time in the context of spin-induced scalarization how a coupling to the Ricci scalar affects all of the above.

In the absence of a Ricci coupling, we find substantial deviations from the Kerr metric in both physical properties and observables. For moderate values of the Ricci coupling, this behaviour persists, but for larger values of this coupling deviations become suppressed significantly. The scalar charge tends to be a good probe for deviations in other observables. Our results are in line with the results of \cite{Antoniou:2021zoy} for curvature induced scalarization and the effect of the Ricci coupling on the charge.

The Ricci coupling is expected to be present from an effective field theory perspective as a lower-order coupling to curvature (in both derivatives and mass dimensions). It is beneficial from a phenomenological perspective as well, by making the scalarization compatible with cosmology \cite{Antoniou:2020nax} and by improving stability \cite{Antoniou:2022agj} and hyperbolicity \cite{Thaalba:2023fmq}. It also quenches scalarization of neutron stars \cite{Ventagli:2021ubn}, which would lead to strong constraints. Only moderate values of the Ricci coupling are needed for all of the above in models that lead to spin-induced scalarization of black holes of a few Solar masses. However, spin-induced scalarization of supermassive black holes requires orders of magnitude larger Gauss-Bonnet coupling. This would in turn source strong tachyonic instabilities for neutron stars and set a minimum mass for black holes above the LIGO/VIRGO/KAGRA mass range. A large Ricci coupling could potentially mitigate these issues, but it would also strongly suppress deviations from Kerr for all observables.

Based on the above, in observationally viable scenarios involving spin-induced scalarization from a quadratic Gauss-Bonnet coupling, we anticipate that any shadow deviations of scalarized black holes from the Kerr metric are unlikely to be discernible in current and future VLBI observations of supermassive black holes.
A potential way out might be models in which supermassive black holes are selectively scalarized, while solar mass black holes and neutron stars are not. Such a model was recently presented in Ref.~\cite{Eichhorn:2023iab} for curvature-induced scalarization, and it would be interesting to study whether similar models exist for the spin-induced case as well.
It would also be interesting to understand whether the addition of further scalar-curvature couplings and/or scalar potentials -- potentially in a multi-field extension of Horndeski gravity -- could alter these findings and allow us to suppress scalarization in neutron stars, while not suppressing the scalar charge of scalarized black holes.

\section*{Acknowledgments}
\noindent P.~F. acknowledges support by a Research Leadership Award from the Leverhulme Trust and A.~E.~and P.~F.~are supported by a grant from Villum Fonden under Grant No.~29405. C.~B. acknowledges partial support from STFC Consolidated Grants no. ST/T000732/1 and no. ST/X000672/1. T.~P.~S. acknowledges partial support from STFC Consolidated Grants no. ST/V005596/1, no. ST/T000732/1 and no. ST/X000672/1.

\bibliography{biblio}

\end{document}